\documentclass[11pt]{article}

\usepackage[final]{acl}

\usepackage{times}
\usepackage{latexsym}

\usepackage[T1]{fontenc}

\usepackage[utf8]{inputenc}

\usepackage{microtype}

\usepackage{inconsolata}
\usepackage{graphicx}
\usepackage{amsmath}
\usepackage{amssymb}
\usepackage{xcolor}
\usepackage{hyperref}
\title{CogErgLLM: Exploring Large Language Model Systems Design Perspective Using Cognitive Ergonomics}

\author{%
    Azmine Toushik Wasi\textsuperscript{1*}, Mst Rafia Islam\textsuperscript{2}\\
\textsuperscript{1}Shahjalal University of Science and Technology, Sylhet, Bangladesh\\
  \textsuperscript{2}Independent University, Dhaka, Bangladesh\\
  \texttt{azmine32@student.sust.edu, 2030391@iub.edu.bd} \\
   \textsuperscript{*}Corresponding author
}

\begin{document}
\maketitle
\begin{abstract}
Integrating cognitive ergonomics with LLMs is crucial for improving safety, reliability, and user satisfaction in human-AI interactions. Current LLM designs often lack this integration, resulting in systems that may not fully align with human cognitive capabilities and limitations. This oversight exacerbates biases in LLM outputs and leads to suboptimal user experiences due to inconsistent application of user-centered design principles. Researchers are increasingly leveraging NLP, particularly LLMs, to model and understand human behavior across social sciences, psychology, psychiatry, health, and neuroscience. Our position paper explores the need to integrate cognitive ergonomics into LLM design, providing a comprehensive framework and practical guidelines for ethical development. By addressing these challenges, we aim to advance safer, more reliable, and ethically sound human-AI interactions.
\end{abstract}



\section{Introduction} \label{sec:intro}
\textit{Ergonomics} focuses on optimizing human-machine interactions for efficiency, safety, and well-being, incorporating both physical and cognitive aspects \cite{Arkouli2022}. \textit{Cognitive science} studies mental processes and behaviours, offering insights crucial for ensuring the safety and reliability of Large Language Models (LLMs) \cite{Qu2024warq, Bermúdez_2020}. 
\textit{Cognitive ergonomics} is the study of how to design systems and interfaces that align with human cognitive abilities and limitations to enhance efficiency, safety, and user satisfaction. It focuses on optimizing mental processes like memory, attention, mental workload and decision-making in human-machine interactions \cite{Bidanda2023-ci}.  
LLMs can also be influenced by various psychological, technological, and decision-specific factors, such as time pressure, emotions, and decision-making styles \cite{eigner2024determinants}, to adapt to human needs and functions.
This convergence with AI supports effective human decision-making, with LLMs designed to enhance transparency and trust, ultimately advancing human-AI interaction systems \cite{LeGuillou2022}.

\begin{figure}[t!] 
\centering {\includegraphics[scale=.2]{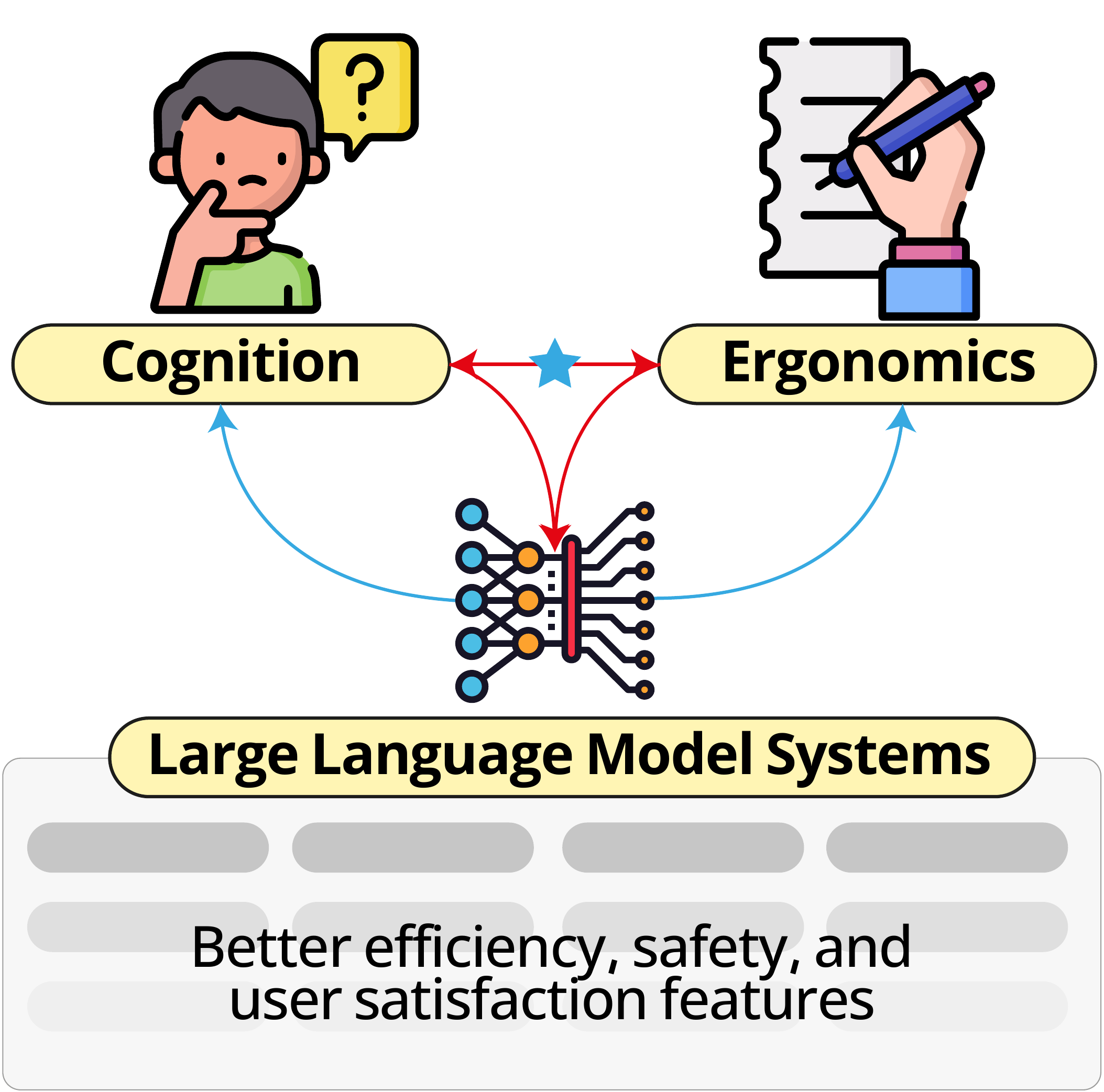}}
\caption{Integration of Cognitive ergonomics and Large Language Models}\label{fig:WHAT}
\end{figure}


Though researchers are actively working in this area, current LLM system design often lacks comprehensive integration of cognitive ergonomics, resulting in systems that may not fully align with human cognitive capabilities and limitations. Secondly, there is insufficient focus on incorporating cognitive science methods to systematically identify and mitigate biases in LLM outputs. Additionally, existing LLM interfaces frequently fail to apply user-centered design principles consistently, leading to sub-optimal user experiences and cognitive overload. Lastly, LLMs often lack mechanisms to explain their decisions and outputs clearly, reducing user trust and transparency, while there's also a noticeable gap in developing LLMs that can adapt to individual user preferences and learning styles over time, hindering their effectiveness and engagement \cite{eigner2024determinants,Subramonyam2024,10.1145/3613905.3650756,LeGuillou2022,wasi2024llmswritingassistantsexploring,wasi2024inkindividualitycraftingpersonalised}. 

In this position paper, we undertake a thorough exploration of the crucial integration of cognitive ergonomics principles into the design framework of LLM systems to address these issues. 
Motivated by the importance of aligning LLM functionalities with human cognitive processes and addressing biases through cognitive science methodologies, we explore the cognitive challenges inherent in LLM designs, proposing a comprehensive design framework grounded in cognitive ergonomics principles, and providing practical guidelines for ethical LLM development. The core contributions of our paper lie in offering a detailed analysis of cognitive ergonomics relevance, outlining a comprehensive design framework,  and recommending future research directions. Through these contributions, we aim to advance understanding and practice in integrating cognitive ergonomics into LLM systems, ultimately fostering safer, more reliable, and ethically sound human-AI interactions.

Our contribution can be summarized in four folds:
\begin{itemize}
\item \textbf{Comprehensive Discussion on Cognitive Ergonomics}: By \textit{CogErgLLM}, we provide a thorough exploration of how cognitive ergonomics principles can be integrated into the design of LLMs. By addressing the gap between LLM design and human cognitive capabilities, we aim to enhance the safety, reliability, and user satisfaction in human-AI interactions.

\item \textbf{Novel Design Framework}: Our paper presents a comprehensive design framework \textit{CogErgLLM}, grounded in cognitive ergonomics principles. This framework offers practical guidelines for ethical LLM development, addressing issues such as bias mitigation, user-centered design, and transparency.

\item \textbf{Case Studies}: We guide the practical application of cognitive ergonomics principles through the development of prototypes and case studies in diverse domains such as healthcare and education, showcasing the effectiveness and versatility of our work.

\item \textbf{Discussion on Challenges and Ethical Considerations}: We identify and discuss technical hurdles and ethical concerns in merging cognitive ergonomics with LLMs, emphasizing the need for continued interdisciplinary research to foster responsible AI development and improve human-AI interaction.
\end{itemize}

\section{Background} \label{sec:bg}
\subsection{Cognitive Ergonomics}
Cognitive ergonomics principles emphasize efficiency, attention support, learning facilitation, decision-making aid, and performance enhancement in interface and system design \cite{asgyhwayhgwayhw2017}. These principles find critical applications in safety-critical environments like air traffic control and medical settings, as well as in everyday domains such as banking and leisure activities, showcasing their broad utility. By leveraging cognitive science knowledge on perception, memory, and problem-solving, cognitive ergonomics aims to optimize human performance and well-being in complex and changeable work environments, ultimately improving productivity and safety while recognizing the importance of human adaptation and the need for adaptable work conditions \cite{Branaghan2020,Parasuraman2003,Dehais2020}.

\subsection{Large Language Models and Cognition}
Recent research has delved into the relationship between LLMs and human cognition, revealing promising insights. Studies by \citet{huff2024psychology} demonstrate LLMs' ability to predict human performance in language-based memory tasks, despite differing internal mechanisms. \citet{shani-etal-2023-towards} explored the development of concept-aware LLMs, showing improved alignment with human intuition and prediction robustness. Additionally, \citet{Samwald2023} compiled core principles for steering and evaluating LLM reasoning, drawn from diverse fields like structured reasoning and ethical guidelines. These advancements underscore LLMs' potential to offer valuable insights into human cognition while emphasizing the importance of safe and effective deployment through rigorous evaluation methods.

\section{Conceptual Foundations} \label{sec:ConceptualFoundations}
Cognitive processes are profoundly influenced by ergonomic design, and aligning these principles with LLMs can enhance their usability, effectiveness, and user satisfaction. Here we describe how it can be done, as described in Figure \ref{fig:ConceptualFoundations}:

\begin{figure*}[t!] 
\centering {\includegraphics[scale=.6]{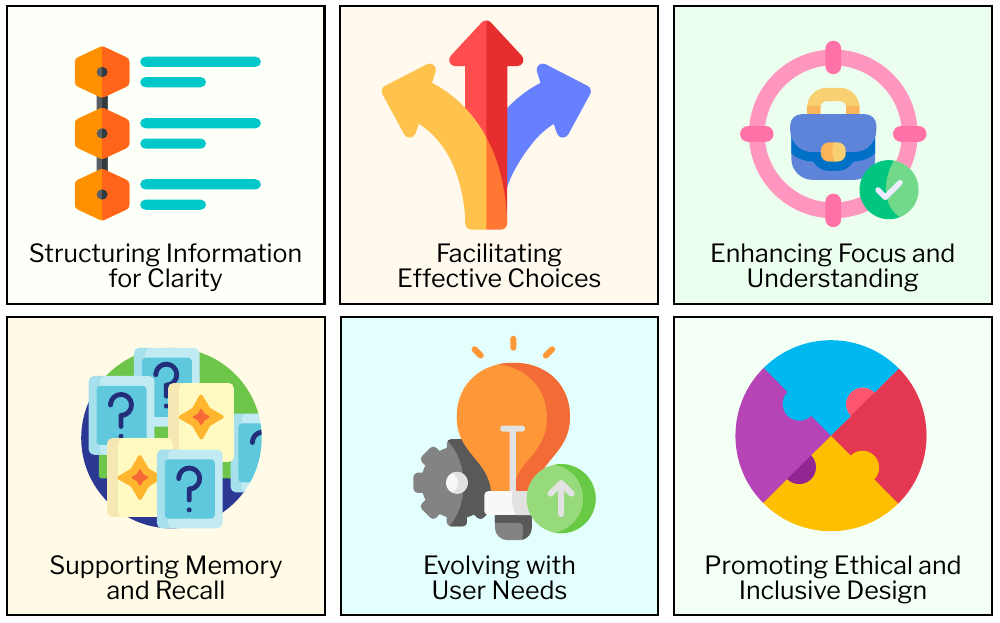}}
\caption{Conceptual Foundations for Cognitive Ergonomics in Large Language Models}\label{fig:ConceptualFoundations}
\end{figure*}

1. \textbf{\textit{Structuring Information for Clarity}: }
   Cognitive ergonomics ensures information is structured clearly and intuitively, reducing cognitive load \cite{SWELLER2024102423}. It focuses on organizing information in a way that reduces mental effort. This can be seen in the design of user interfaces, where clear layouts and intuitive navigation systems help users quickly find what they need. For instance, websites often use hierarchical menus and clear categorization to enhance usability, ensuring visitors can navigate without confusion.
   
2. \textbf{\textit{Facilitating Effective Choices}:}
   Ergonomic design aids decision-making by providing essential information clearly and minimizing cognitive biases \cite{articfewtgle}. For example, in retail environments, ergonomic principles guide the placement of products to highlight choices effectively, helping customers make informed decisions. This approach reduces cognitive load and ensures that decisions are based on relevant information rather than misleading cues.
   
3. \textbf{\textit{Enhancing Focus and Understanding}:}
   Ergonomic design focuses on how users perceive and attend to information, placing important elements where they are easily noticed \cite{doi:https://doi.org/10.1002/9781119636113.ch3}.  For example, in classrooms, teachers use visual aids and highlighted content on slides to direct students' attention to key concepts. This practice ensures that important information is conveyed effectively, promoting better understanding and retention among learners.

4. \textbf{\textit{Supporting Memory and Recall}:}
   Strategies like familiar patterns and clear information presentation are crucial in ergonomic design to ease the burden on users' memory \cite{doi:https://doi.org/10.1002/9781119636113.ch5}. Such as, in museum exhibits, interactive displays often repeat key information and use consistent labeling to reinforce learning. This approach helps visitors retain information and recall it later, enhancing their overall experience and educational outcomes.

5. \textbf{\textit{Evolving with User Needs}:}
   Cognitive ergonomics emphasizes the importance of adapting designs based on user feedback and preferences \cite{doi:https://doi.org/10.1002/9781119636113.ch41}. In software development, agile methodologies allow teams to iterate quickly based on user testing and input. This flexibility ensures that products evolve to meet changing user needs, enhancing satisfaction and usability over time.
   
6. \textbf{\textit{Promoting Ethical and Inclusive Design}:}
Incorporating cognitive ergonomics into LLM design can promote ethical and inclusive practices. By considering diverse user needs and cultural backgrounds during the design process, LLMs can be developed to be more accessible and equitable \cite{doi:https://doi.org/10.1002/9781119636113.ch9}. 
For example, incorporating cognitive ergonomics into urban planning promotes ethical and inclusive design by prioritizing accessibility and user-centered approaches. This involves integrating features like ramps, tactile paving, and inclusive seating to ensure public spaces accommodate individuals with disabilities. Engaging with diverse communities throughout the design process ensures their needs are met, fostering a sense of inclusivity and social equity. Ethical considerations guide these efforts, emphasizing dignity and equal access to essential services. Such practices not only enhance physical accessibility but also promote social integration and community well-being, establishing cities as advocates for ethical and inclusive urban development.

As discussed above, integrating principles of cognitive ergonomics into website design offers several significant benefits. It enhances usability by simplifying navigation and minimizing the learning curve for visitors. By organizing information clearly, it reduces cognitive load and supports effective decision-making by presenting options and data in a straightforward manner. Enhancing transparency in data handling and user control over privacy settings increases trust among users. Personalizing content delivery based on user preferences and behavior not only improves relevance but also enhances user engagement and satisfaction. Moreover, implementing strategies to mitigate biases in content presentation ensures fairness and ethical standards. Overall, integrating cognitive ergonomics in web design leads to improved user experience, making websites more efficient, user-friendly, and trustworthy platforms for information and interaction.


\section{Design Framework} \label{sec:DesignFramework}
In this section, we describe the main components of CogErgLLM, a framework designed to explore and integrate cognitive ergonomics principles into the development of LLMs. The framework aims to enhance user experience, efficiency, and reliability by aligning LLM interactions with human cognitive processes. 

\subsection{Methodology}
The development methodology of the framework components, draws inspiration from theoretical foundations outlined in Sections \ref{sec:intro} and \ref{sec:bg}. These theoretical underpinnings of cognitive ergonomics serve as guiding principles for the design and definition of each framework element. We meticulously define and conceptualize the various components, informed by theories on human cognition and interaction with technology. Then, we transition to the implementation phase, where we translate the design aspects into tangible prototypes and undergo rigorous evaluation processes to assess their efficacy in enhancing user experience and system performance.

\subsection{Components of CogErgLLM}
\begin{figure*}[t!] 
\centering {\includegraphics[scale=.7]{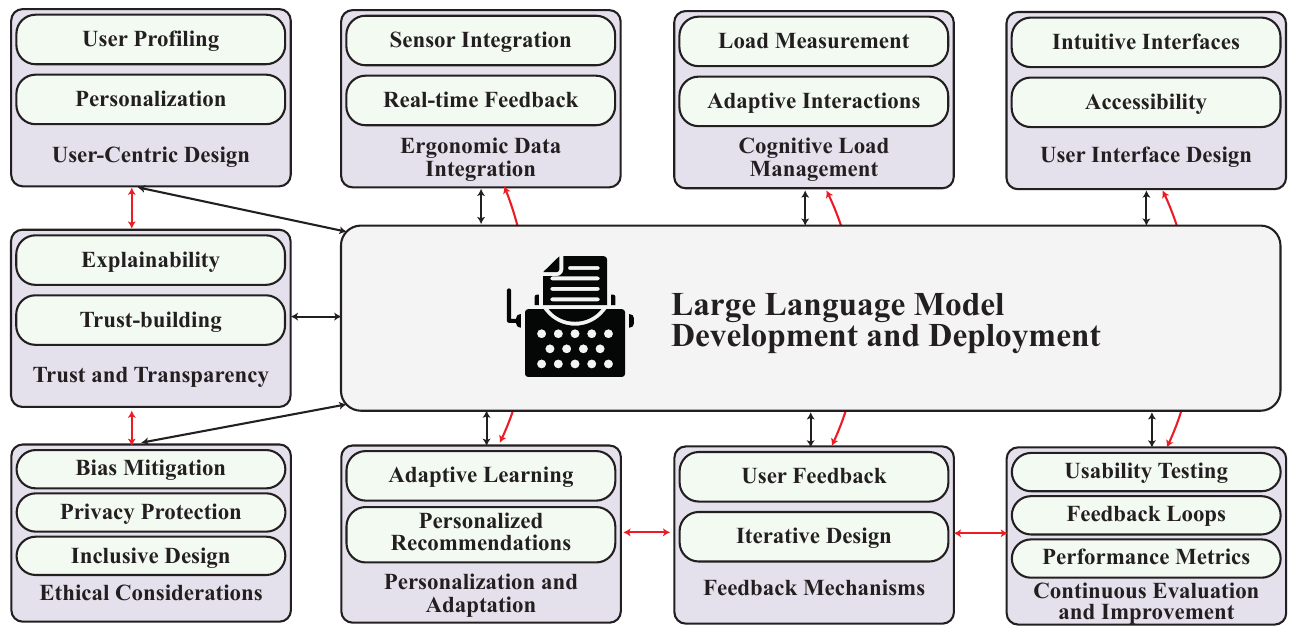}}
\caption{Components of CogErgLLM}\label{fig:Components}
\end{figure*}

We outline key components such as user-centric design, ergonomic data integration, cognitive load management, user interface design, trust and transparency, feedback mechanisms, and more, as described in Figure \ref{fig:Components}. Each component is crucial for creating LLM systems that are intuitive, adaptive, and supportive of users' cognitive needs.

\subsubsection{User-Centric Design}
\textbf{User Profiling}: Understanding user needs and preferences is crucial for tailoring LLM interactions to individual users. Techniques such as surveys, interviews, and behavioral analysis can provide insights into users' cognitive capabilities and limitations, ensuring that LLMs are designed to meet diverse requirements effectively. By incorporating these profiles, LLMs can adapt their responses to be more relevant and engaging, enhancing the overall user experience \cite{wang2024llmenhanceduseriteminteractionsleveraging}.

\noindent \textbf{Personalization}: Personalizing LLM interactions based on user profiles can significantly improve usability and satisfaction \cite{li2024personalllmagentsinsights}. Cognitive ergonomics emphasizes the importance of designing systems that align with users' mental models and preferences. By using data from user profiles, LLMs can offer tailored suggestions, responses, and content, making interactions more intuitive and reducing cognitive strain \cite{wasi2024inkindividualitycraftingpersonalised}.

\subsubsection{Ergonomic Data Integration}

\textbf{Sensor Integration}: Incorporating ergonomic sensors to monitor user posture and environment can provide valuable data for optimizing LLM interactions \cite{Luo20243rwqtwq3t, xu2024generalpurposedeviceinteractionllms}. For example, sensors can detect when a user is experiencing physical discomfort or cognitive fatigue, prompting the LLM to adjust its interaction style or offer breaks. This integration helps in creating a more comfortable and supportive user environment.

\noindent \textbf{Real-time Feedback}: Providing real-time ergonomic advice based on sensor data can enhance user well-being and productivity \cite{xu2024penetrativeaimakingllms,Luo20243rwqtwq3t}. For instance, if sensors detect that a user has been sitting in a poor posture for an extended period, the LLM can offer corrective suggestions. This immediate feedback loop ensures that users maintain optimal ergonomic conditions, reducing physical and cognitive stress.

\subsubsection{Cognitive Load Management}

\textbf{Load Measurement}: Tools and methods for assessing cognitive load, such as eye-tracking and brainwave analysis, can help designers understand how users interact with LLMs. This data is crucial for identifying points of high cognitive load and areas where the system may be causing unnecessary strain, allowing for targeted improvements \cite{Krell202wg4gt4w2}.

\noindent \textbf{Adaptive Interactions}: Strategies for adjusting LLM interactions to manage cognitive load include simplifying complex information, providing information progressively, and offering clear, concise instructions. By adapting to the user's cognitive state, LLMs can ensure that interactions remain manageable and effective, preventing overload and enhancing comprehension \cite{subramonyam2024bridginggulfenvisioningcognitive}.

\subsubsection{User Interface Design}

\textbf{Intuitive Interfaces}: Design principles for creating user-friendly LLM interfaces focus on minimizing unnecessary complexity and enhancing navigability. Cognitive ergonomics principles such as consistency, predictability, and immediate feedback ensure that users can easily understand and interact with the system, improving efficiency and satisfaction.

\noindent \textbf{Accessibility}: Ensuring interfaces are accessible to all users, including those with disabilities, is a key aspect of ergonomic design. This includes adhering to standards such as WCAG and incorporating features like text-to-speech, adjustable font sizes, and high-contrast modes. Accessibility ensures that LLMs are usable by a broader audience, promoting inclusivity and equity.

\subsubsection{Trust and Transparency}
\noindent 
\textbf{Explainability}: Techniques for making LLM decisions transparent to users include providing clear, understandable explanations for actions and recommendations. This transparency helps users understand how the LLM works, fostering trust and confidence in the system. Cognitive ergonomics emphasizes the need for systems that users can predict and rely on.

\noindent 
\textbf{Trust-building}: Strategies for enhancing user trust in LLMs involve consistent, reliable performance, and the ability to demonstrate ethical and unbiased behavior \cite{2024exploringbengalireligious}. Incorporating user feedback and continuously improving the system based on that feedback also plays a crucial role in building and maintaining trust.

\subsubsection{Feedback Mechanisms}
\noindent 
\textbf{User Feedback}: Incorporating user feedback into the design process allows for continuous improvement of LLM performance. Cognitive ergonomics highlights the importance of listening to users and adapting systems to meet their evolving needs. Regularly gathering and analyzing feedback helps identify areas for enhancement and ensures that the system remains aligned with user expectations \cite{pan2024feedbackloopslanguagemodels,xu2024generalpurposedeviceinteractionllms}.

\noindent 
\textbf{Iterative Design}: Employing iterative design processes based on feedback ensures that LLMs are continually refined and optimized \cite{pan2024feedbackloopslanguagemodels}. This approach allows for the gradual incorporation of new ergonomic insights and user requirements, leading to more effective, user-friendly, and reliable systems. Iterative design helps in addressing issues promptly and evolving the system to better serve its users.

\subsubsection{Ethical Considerations}
\noindent 
\textbf{Bias Mitigation}: Incorporating cognitive ergonomics into LLM design involves creating processes that actively identify and mitigate biases in outputs. By leveraging cognitive science methods, such as social intelligence tests and moral dilemma scenarios, designers can ensure that LLMs provide fair and equitable responses, thus aligning with diverse user expectations and promoting ethical AI use.

\noindent 
\textbf{Privacy Protection}: Robust data security and privacy measures are essential in designing LLMs that respect user confidentiality. Cognitive ergonomics emphasizes understanding users' cognitive concerns about data security, thus guiding the implementation of intuitive privacy controls and transparent data usage policies that reassure users and protect their personal information.

\noindent 
\textbf{Inclusive Design}: Ensuring that LLMs cater to diverse user groups requires an inclusive design approach informed by cognitive ergonomics. This involves creating interfaces and interactions that consider the cognitive and cultural backgrounds of different users, making the systems accessible and usable for people with varying linguistic needs and cognitive abilities, thereby promoting inclusivity and diversity.

\subsubsection{Personalization and Adaptation}

\textbf{Adaptive Learning}: Integrating cognitive ergonomics principles into LLM design enables the development of systems that dynamically adapt to individual user preferences and learning styles. By leveraging cognitive science insights, such as theories of learning and memory, designers can create LLMs that personalize content delivery and interactions based on users' behavior and performance, thereby enhancing learning outcomes and user satisfaction.

\noindent \textbf{Personalized Recommendation}: Cognitive ergonomics informs the implementation of personalized recommendation systems within LLMs, leveraging user data to offer tailored content and suggestions. By analyzing user interactions and preferences, designers can optimize content relevance and engagement, enhancing user experience and promoting continued usage of the LLMs. Additionally, incorporating principles of cognitive load management ensures that recommendations are presented in a manner that optimizes cognitive resources and minimizes information overload.

\subsubsection{Continuous Evaluation and Improvement}

\textbf{Usability Testing}: Cognitive ergonomics emphasizes the importance of usability testing in LLM development, involving real users to identify usability issues and areas for improvement. By conducting regular usability tests, designers can gather valuable feedback on user interactions, navigation patterns, and comprehension levels, enabling iterative refinement of LLM interfaces and functionalities to better align with users' cognitive needs and preferences.

\noindent \textbf{Feedback Loops}: Establishing feedback mechanisms is integral to the continuous improvement of LLM interactions \cite{pan2024feedbackloopslanguagemodels}. By collecting and analyzing user input and suggestions, designers can gain insights into user satisfaction, comprehension difficulties, and feature preferences, allowing for timely adjustments and enhancements to LLM systems. Feedback loops ensure that LLMs remain responsive to evolving user requirements and cognitive dynamics, promoting sustained user engagement and system effectiveness.

\noindent \textbf{Performance Metrics}: Defining and monitoring key performance indicators (KPIs) related to cognitive ergonomics \cite{Zhang2019wafaqf} provides valuable insights into the effectiveness of LLM designs. Metrics such as user satisfaction, task completion rates, and cognitive load assessments enable designers to evaluate the impact of ergonomic interventions on user experience and system performance, guiding optimization efforts and ensuring the continued delivery of user-centric LLM solutions.

\section{Case Studies}
Use case studies demonstrate the practical application of the CogErgLLM framework in real-world scenarios. In healthcare, the framework supports medical professionals by presenting critical patient information and treatment options clearly and concisely. In education, it tailors learning experiences to individual students, enhancing engagement and knowledge retention through adaptive learning and memory support techniques.

\textbf{Healthcare.}
In a healthcare setting, CogErgLLM can be used to assist medical professionals in making informed decisions. For instance, the system integrates with electronic health records (EHR) to provide doctors with a summary of patient histories, relevant medical literature, and potential treatment options. The LLM uses cognitive load management techniques to present this information in manageable chunks, reducing the mental effort required by physicians. Adaptive learning algorithms personalize the interface based on individual doctors' specialities and preferences, enhancing usability and efficiency. Real-time feedback mechanisms alert medical staff to potential issues, ensuring quick and accurate decision-making. By incorporating cognitive ergonomic principles, this application aims to improve patient outcomes, reduce cognitive fatigue among healthcare providers, and streamline clinical workflows.

\textbf{Education.} 
In the educational sector, CogErgLLM will serve as a personalized learning assistant for students. The system tailors content delivery to match each student's learning style and pace, using adaptive learning technologies. For example, it breaks down complex topics into simpler subtopics, gradually increasing complexity as the student's understanding improves. Memory considerations are addressed by incorporating spaced repetition techniques and interactive quizzes to reinforce learning. The interface supports multimodal interactions, allowing students to engage through text, voice, and visual aids. Teachers receive insights into student progress and cognitive load, enabling them to adjust their teaching strategies accordingly. This application of cognitive ergonomics aims to enhance student engagement, improve knowledge retention, and provide a more personalized and effective learning experience.

\textbf{Legal Work.}  
CogErgLLM can significantly enhance efficiency and accuracy in legal work by providing advanced support to legal professionals. For example, in a case involving complex contract review, CogErgLLM can integrate with legal databases and case management systems to assist lawyers in analyzing and summarizing extensive legal documents. The system uses cognitive ergonomics principles to break down lengthy contracts into more digestible sections, highlighting key clauses and potential issues. It employs natural language processing to compare the contract against relevant case law and statutes, offering suggestions for amendments and flagging areas of concern. Adaptive learning algorithms customize the interface based on the lawyer's specialization and previous cases, improving relevance and usability. Real-time feedback mechanisms alert lawyers to critical deadlines and compliance requirements. By reducing cognitive load and streamlining document review processes, CogErgLLM aims to enhance legal research efficiency, improve accuracy in contract analysis, and support lawyers in delivering more informed and timely legal advice.

\textbf{Creative Writing.}  
CogErgLLM can transform the creative writing process by offering tailored assistance and enhancing the writer's productivity. For instance, in the case of drafting a novel, CogErgLLM can integrate with writing tools and databases to provide real-time support. The system analyzes the writer's style and narrative structure, suggesting plot developments, character traits, and dialogue options that align with the writer's creative vision. It employs cognitive load management techniques to break down complex narrative arcs into manageable segments, offering feedback and guidance on pacing and coherence. By using adaptive learning algorithms, CogErgLLM personalizes its suggestions based on the writer's genre preferences and past works, ensuring relevance and enhancing creativity. The system also includes interactive features, such as brainstorming prompts and scenario simulations, to inspire new ideas and overcome writer's block. Through real-time feedback and contextual assistance, CogErgLLM aims to streamline the creative process, reduce cognitive fatigue, and support writers in crafting compelling and cohesive stories.

\textbf{Emergency Response System.}
Implementing CogErgLLM can revolutionize our city's emergency response system by integrating advanced natural language understanding and cognitive ergonomics. This intelligent system assists dispatchers and first responders by swiftly analyzing emergency calls, prioritizing incidents, and recommending optimal response strategies in real-time. As a result, we've significantly reduced average response times, enhanced dispatcher efficiency, and improved accuracy in incident classification. The intuitive interface and actionable insights provided by CogErgLLM have garnered positive feedback from emergency personnel, paving the way for future enhancements such as predictive analytics and real-time data integration to further elevate our emergency management capabilities.

\section{Discussion}
We believe that the integration of cognitive ergonomics with LLMs represents a significant step forward in enhancing the usability, effectiveness, and ethical integrity of AI systems. By applying cognitive ergonomic principles to LLM design, we can create interfaces and interactions that are more intuitive, transparent, and aligned with human cognitive capabilities. This not only improves user experience but also fosters trust and collaboration between humans and AI. However, this integration is not without its challenges, particularly in addressing technical complexities, ensuring data privacy, and mitigating biases. Despite these hurdles, we see immense potential in the future of cognitive ergonomics in LLMs, offering opportunities for innovative research, inclusive design practices, and the advancement of human-AI interaction. Through concerted efforts and interdisciplinary collaboration, we can harness the power of cognitive ergonomics to shape a future where AI systems truly augment human capabilities while upholding ethical standards and promoting user well-being.

\noindent
\textbf{Cognitive Ergonomics with LLMs for Industrial Applications}.
Cognitive ergonomics in industrial setups, particularly when applied to large language models (LLMs), offers promising potential for improving user interactions and operational efficiency. By integrating cognitive ergonomics principles, LLMs can be designed to align more closely with human cognitive processes, leading to more intuitive and effective interfaces. This can enhance task automation, decision support, and human-computer collaboration within industrial environments. However, implementing these principles presents challenges, including the need for context-specific adaptation and the integration of complex cognitive models into LLM systems. Future research should focus on developing practical frameworks and tools for applying cognitive ergonomics in industrial contexts, evaluating their impact on productivity and user satisfaction, and addressing technical limitations such as model scalability and data privacy. Such advancements could significantly enhance the usability and effectiveness of LLMs in various industrial applications.

\noindent
\textbf{Technical Challenges}.
Integrating cognitive ergonomics with LLMs presents several technical challenges that need to be addressed. One significant challenge is ensuring that cognitive ergonomic principles can be effectively translated into AI system designs without compromising performance or functionality. This includes developing interfaces that are both intuitive and capable of handling complex user interactions. Another challenge is maintaining data privacy while implementing personalized and adaptive features, as balancing user customization with robust data protection is crucial. Additionally, mitigating biases in LLM outputs while applying cognitive ergonomics requires careful consideration to ensure fairness and equity. Technical complexities also arise in adapting cognitive ergonomic principles to diverse user needs and contexts, necessitating advanced algorithms and continuous refinement. Addressing these challenges requires ongoing research, interdisciplinary collaboration, and innovative solutions to fully integrate cognitive ergonomics into LLMs while upholding technical integrity and ethical standards.

\noindent
\textbf{Future Opportunities}.
Despite the challenges, integrating cognitive ergonomics with LLMs presents numerous future opportunities for research and development. One avenue for exploration is the enhancement of LLM interpretability and explainability through cognitive ergonomic design, enabling users to better understand and trust LLM outputs. Additionally, leveraging cognitive ergonomics to tailor LLM interactions to diverse user demographics and preferences opens doors for inclusive and personalized AI experiences. Furthermore, exploring novel applications of cognitive ergonomic principles in LLM design, such as emotion recognition and adaptive learning, holds promise for advancing human-AI interaction capabilities.

\vspace{-1mm}
\section{Conclusion}
\vspace{-1mm}
Our paper presents \textit{CogErgLLM}, a framework which integrates cognitive ergonomics principles into the design of LLMs. Our core contributions include the comprehensive exploration and integration of cognitive ergonomics, development of a design framework, practical case studies, and recommendations for future research. It holds significant potential to enhance human-AI interaction by improving safety, reliability, and user satisfaction. Its impact extends across various domains, from healthcare to workplace settings, where LLMs play critical roles. We encourage further research and collaboration in this interdisciplinary area to advance the understanding and implementation of ethical, user-centric AI systems, laying the foundation for a truly human-centric advanced artificial general intelligence.

\clearpage
\newpage

\section*{Limitations}
This study focuses solely on the theoretical perspective of cognitive ergonomics in relation to LLMs and does not extend to the development of mathematical models or empirical evaluations. While it provides a valuable conceptual framework for integrating CE principles with LLM design, it lacks practical contributions such as algorithmic implementation or performance assessment. The absence of quantitative models and evaluative methods limits the study’s ability to directly address how these theoretical principles can be operationalized and tested within LLM systems. Future research should aim to bridge this gap by developing and validating mathematical models and evaluation techniques to apply cognitive ergonomics more effectively in practical LLM applications.

\section*{Ethical Considerations}
Ethical concerns surrounding the integration of cognitive ergonomics with LLMs revolve primarily around data privacy and bias mitigation. With LLMs relying heavily on vast amounts of data for training, ensuring the privacy and security of user data becomes paramount. Striking a balance between collecting sufficient data for effective cognitive ergonomic design and safeguarding user privacy requires robust encryption techniques, anonymization protocols, and transparent data handling practices. Moreover, mitigating biases inherent in LLMs, stemming from the biases present in training data, poses ethical challenges. Addressing biases demands proactive measures such as diverse dataset curation, algorithmic fairness assessments, and continuous monitoring and adjustment of LLMs to minimize discriminatory outcomes.

\section*{Acknowledgements}
We express our sincere gratitude to Computational Intelligence and Operations Laboratory (CIOL) for their support.

\section*{Author Contributions}
ATW contributed to most of the work, including conceptualization, formal analysis, application, and writing. MRI contributed to the formulation of legal aspects and related case studies and writing.


\bibliography{work}

\clearpage
\newpage

\appendix

\end{document}